\documentclass[12pt]{article}
 cm
\usepackage{amsfonts,amssymb}
\usepackage{epsfig, picinpar}

\newtheorem{theorem}{Theorem}

\newtheorem{remark}{Remark}

\newcommand{\ju}[4]{\mbox{$
  \left(\begin{array}{cc}{#1} & {#2}\\{#3} &{#4}  \end{array}\right)$}}

\newcommand{\zhen}[9]{\mbox{$
  \left(\begin{array}{ccc}{#1} & {#2} & {#3}\\{#4} & {#5} & {#6}\\
{#7} &{#8} &{#9}  \end{array}\right)$}}
\newcommand{\plem}[3]{\mbox{$\left(\begin{array}{c} \psi_{#1}\\ \psi_{#2}\\ \psi_{#3}   \end{array} \right)$}}
\newcommand{\phlem}[3]{\mbox{$\left(\begin{array}{c} \psi^*_{#1}\\ \psi^*_{#2}\\ \psi^*_{#3}   \end{array} \right)$}}

\newcommand{\R}{\mathbb{R}}

\newcommand{\Z}{\mathbb{Z}}

\newcommand{\pa}{\partial}

\newcommand{\al}{\alpha}  
\newcommand{\be}{\begin{equation}}
\newcommand{\ee}{\end{equation}}
\newcommand{\beq}{\begin{eqnarray}}
\newcommand{\eeq}{\end{eqnarray}}
\newcommand{\beqq}{\begin{eqnarray*}}
\newcommand{\eeqq}{\end{eqnarray*}}

\newcommand{\La}{\Lambda}

\title{\bf Integrable hierarchy, $3\times 3$ constrained systems, and
  parametric and peaked stationary solutions}
\author{Darryl D. Holm,\ \  Zhijun Qiao\\
T-7 and CNLS, MS B-284, Los Alamos National Laboratory\\
Los Alamos, NM 87545, USA\\
{\footnotesize  E-mails:  dholm@lanl.gov\ \ \ qiao@lanl.gov}
}
\date{First version Jan. 28, 2002\\
Second version July 31, 2002}
\begin{document}
\maketitle

\begin{abstract}
This paper gives a new integrable hierarchy of nonlinear evolution
equations. The DP equation: $m_t+um_x+3mu_x=0, \ m=u-u_{xx}$,
proposed recently by Desgaperis and Procesi \cite{DP[1999]}, 
is the first one in the negative
hierarchy while the first one in the positive hierarchy
is:\
$m_t=4(m^{-\frac{2}{3}})_x-5(m^{-\frac{2}{3}})_{xxx}+
(m^{-\frac{2}{3}})_{xxxxx}$.
The whole hierarchy is shown Lax-integrable through
solving a key matrix equation. To obtain the parametric solutions for
the whole hierarchy, we separatedly discuss the negative and the
positive hierarchies. For the negative hierarchy,
its $3\times3$ Lax pairs
  and
 corresponding adjoint representations 
 are nonlinearized to be 
Liouville-integrable  Hamiltonian canonical systems under the so-called
Dirac-Poisson bracket defined 
on a symplectic submanifold of $\R^{6N} $.
 Based on the integrability
of those finite-dimensional canonical Hamiltonian systems we give
the parametric solutions of the all equations in the 
negative hierarchy. In particular, we obtain
the parametric solution of the DP equation.
Moreover, for the positive hierarchy, 
we consider the different constraint and use
a similar procudure to the negative case
to obtain 
the parametric solutions of the
positive hierarchy. In particular, we give 
the parametric solution of the 5th-order PDE 
$m_t=4(m^{-\frac{2}{3}})_x-5(m^{-\frac{2}{3}})_{xxx}+
(m^{-\frac{2}{3}})_{xxxxx}$.
Finally, we discuss the stationary solutions of  the 5th-order PDE,
and particularly give its four  peaked stationary solutions.
The stationary solutions  may be included in the
parametric solution, but the peaked stationary solutions
not. The  5th-order PDE does not have the cusp soliton although
it looks like a higher order Harry-Dym equation.

\end{abstract}
   
   $ $\\
{\it AMS Subject: 35Q53; 58F07; 35Q35}

\section{Introduction}

The inverse scattering transformation (IST) method is a powerful tool
to solve  integrable nonlinear evolution equations (NLEEs) \cite{GGKM}.
  This method has been successfully applied
to give soliton solutions of the integrable NLEEs.
These examples include the well-known KdV
equation \cite{KDV}, which is related to a 2nd order operator
(i.e. Hill operator) 
spectral problem \cite{LG, Mar}, the remarkable AKNS equations
\cite{AKNS, AKNS1}, which is associated with the Zakharov-Shabat (ZS)
spectral problem \cite{ZS}, and other higher dimensional
integrable equations.

To look for as many 
integrable systems as possible has been an important topic
In the theory of
integrable system.
 Kaup \cite{Kaup1} studied
the inverse scattering problem for cubic eigenvalue equations of the
form $\psi_{xxx}+6Q\psi_x+6R\psi=\lambda\psi$, and showed a 5th
order partial differential equation (PDE)
 $Q_t+Q_{xxxxx}+30(Q_{xxx}Q+\frac{5}{2}Q_{xx}Q_x)
+180Q_xQ^2=0$ (called the KK equation) integrable. Afterwards, Kuperschmidt \cite{Kup1}
constructed a super-KdV equation and presented the integrability of
the equation through giving bi-Hamiltonian property and Lax
form. In 1984  Konopelchenko and Dubrovsky \cite{KD1}
 presented a 5th order equation $u_t=(u^{-2/3})_{xxxxx}$ and pointed out
that this equation is a reduction of some $2+1$ dimensional
equation. We have found the parametric solution 
and some traveling wave solutions of this equation \cite{HQ-new-twv}.

Recently, Degasperis and Procesi \cite{DP[1999]}
proposed a new integrable equation: $m_t+um_x+3mu_x=0, \ m=u-u_{xx} $,
called the DP equation, which has   the peaked soliton solution.  The DP equation is an extention of the Camassa-Holm (CH) equation
 \cite{CH1},  and is proven to be associated with
a 3rd order spectral problem \cite{DHH2002}:
$\psi_{xxx}=\psi_x-\lambda m\psi $ and to have some
relationship to a canonical Hamiltonian system
under a new nonlinear  Poisson bracket (called Peakon Bracket) 
 \cite{HH[2002]}.

The purpose of the present paper has two folds:
\begin{itemize}
\item extend the DP equation to a new  integrable hierarchy of NLEEs
  through studying the functional gradient of the spectral problem and
  a pair of Lenard's operators;
\item
 connect the DP equation to some finite-dimensional
integrable system,  give its parametric solution
from the view point of constraint, and furthermore 
study the parametric solution,
stationary solution and cusp soliton solution.
\end{itemize}

The whole paper is organized as follows.
Next section is saying how to connect a spectral problem to the DP 
equation  and how to cast it into a new hierarchy of
NLEEs, and is also giving the pair of Lenards operators
for the whole hierarchy. In section 3, we construct
the zero curvature representations for this new hierarchy
through solving a key $3\times 3 $ matrix equation, and therefore this
hierarchy 
is integrable. In particular, we
give the matrix Lax pair of the DP equation, which  is equivalent to
the form in Ref. \cite{DHH2002}, as well as  the Lax pair for a 5th-order equation $
m_t=4(m^{-\frac{2}{3}})_x-5(m^{-\frac{2}{3}})_{xxx}+
(m^{-\frac{2}{3}})_{xxxxx}$. We will see that the DP equation is included in the
negative hierarchy while the 5th-order equation in the positive
hierarchy. 
   To obtain the parametric solutions for
the whole hierarchy, we separatedly discuss the negative and the
positive hierarchies. In section 4, we deal with the negative hierarchy.
Its $3\times3$ Lax pairs
  and
 corresponding adjoint representations 
 are nonlinearized to be 
Liouville-integrable  Hamiltonian canonical systems under the so-called
Dirac-Poisson bracket defined 
on a symplectic submanifold of $\R^{6N} $.
 Based on the integrability
of those finite-dimensional canonical Hamiltonian systems we give
the parametric solutions of the all equations in the 
negative hierarchy. In particular, we obtain
the parametric solution of the DP equation.
Section 5  copes with the positive hierarchy. 
We consider the different constraint between the
potential and the eigenfunctions. Under this constraint
the $3\times3$ Lax pairs
  and
 corresponding adjoint representations of the positive hierarchy  are nonlinearized to be 
Liouville-integrable  Hamiltonian canonical systems
in the whole $ \R^{6N}$. Then we obtain 
the parametric solutions of the
positive hierarchy. In particular, we give 
the parametric solution of the 5th-order PDE 
$m_t=4(m^{-\frac{2}{3}})_x-5(m^{-\frac{2}{3}})_{xxx}+
(m^{-\frac{2}{3}})_{xxxxx}$.
Finally, in section 6 we discuss the stationary solutions of  the 5th-order PDE,
and particularly give its four  peaked stationary solutions.
The stationary solutions  may be included in the
parametric solution, but the peaked stationary solutions
not. The 5th-order PDE does not have the cusp soliton, either.

\section{Spectral problems and Lenards operators}
Let us consider the following spectral problem
\be
\psi_{xxx}=\frac{1}{\al^2}\psi_x-\lambda m\psi \label{sp1}
\ee
and its adjoint problem
\be
\psi^*_{xxx}=\frac{1}{\al^2}\psi^*_x+\lambda m\psi^*, \label{sp2}
\ee
where $\lambda $ is a spectral parameter, $m$ is a scalar potential
function, $ \psi$ and $ \psi^*$ are the spectral wave functions
corresponding to the same
$\lambda $, $ \psi^*$ is not conjugate of $ \psi$, and $\al=constant $. 
Then, we have
\be
\nabla\lambda=\frac{\delta\lambda}{\delta m}=\frac{\lambda\psi\psi^*}{E} \label{fg1}
\ee
where
\beqq
E=-\int_{-\infty}^{\infty}m\psi\psi^* dx=const.
\eeqq
Here during our computation about the functional gradient
$\frac{\delta\lambda}{\delta m}$ of the spectral parameter $\lambda$ with respect
to the potential $m$,
 we need the boundary conditions of decaying at infinities
or periodicity condition for the potential function $m$.
A general calculated method can be seen in Refs. \cite{Cao1989,Tu1990}.

Now, we denote the function $\nabla_1\lambda$ by
\be
\nabla_1\lambda=\frac{\lambda(\psi\psi^*_x-\psi^*\psi_x)}{E},   \label{fg2}
\ee
Then, we have the following equality:
\beqq
\bar{K}(\nabla\lambda,\nabla_1\lambda)^T=\lambda \bar{J}(\nabla\lambda,\nabla_1\lambda)^T \label{KJ1}
\eeqq
where $\bar{K} $ and $\bar{J} $ are two matrix operators
\beqq
\bar{K}&=&\ju{-4\pa+5\al^2\pa^3-\al^4\pa^5}{0}{0}{\pa^3-\frac{1}{\al^2}\pa},\label{K1}\\
\bar{J}&=&\ju{0}{3\al^4(2m\pa+\pa m)}{m\pa+2\pa m}{0}\label{J1},\\
\pa&=&\frac{\pa}{\pa x}, \nonumber
\eeqq
or we rewrite Eq. (\ref{KJ1}) as the following Lenard spectral problem form
\beq
K\nabla\lambda&=&\lambda^2J\nabla\lambda, \label{KJ2}
\eeq
where
\beqq
K&=&4\pa-5\al^2\pa^3+\al^4\pa^5,\label{K2}\\
J&=&3\al^6(2m\pa+\pa m)(-\al^2\pa^3+\pa)^{-1}(m\pa+2\pa m). \label{J2}
\eeqq

Without loss of generality, we assume $\al=1$ below.
Then, $K,J$ read
\beq
K&=&4\pa-5\pa^3+\pa^5, \label{K3}\\
J&=&3(2m\pa+\pa m)(\pa-\pa^3)^{-1}(m\pa+2\pa m). \label{J3}
\eeq
 
{\bf Hint: } \ Here we do not care about the Hamiltonian properties
of the operators $K, J$, but need
\beqq
K^{-1}&=&(\pa-\pa^3)^{-1}(4-\pa^2)^{-1},\\
J^{-1}&=&\frac{1}{27}m ^{-2/3}\pa^{-1}m ^{-1/3}(\pa-\pa^3)m ^{-1/3}\pa^{-1}m ^{-2/3}.
\eeqq
They yield 
\beqq
& &{\cal L}=J^{-1}K=\frac{1}{27}m ^{-2/3}\pa^{-1}m ^{-1/3}(\pa-\pa^3)m ^{-1/3}\pa^{-1}m ^{-2/3}(4\pa-5\pa^3+\pa^5),\\
& &{\cal L}^{-1}=K^{-1}J=3(\pa-\pa^3)^{-1}(4-\pa^2)^{-1}(2m\pa+\pa m)(\pa-\pa^3)^{-1}(m\pa+2\pa m),
\eeqq
which are actually the two recursive operators we need in the next section.

\section{Zero curvature representations and an integrable  hierarchy}

Letting $\psi=\psi_1 $, we change Eq. (\ref{sp1}) to a $3\times 3 $ matrix spectral problem
\beq
\Psi_x&=&U(m,\lambda)\Psi,  \label{SP1}\\
U(m,\lambda)&=&\zhen{0}{1}{0}{0}{0}{1}{-m\lambda}{1}{0},\ \
\Psi=\plem{1}{2}{3}. \label{U1}
\eeq

Apparently, the Gateaux derivative matrix  $U_{*}(\xi)$ of the
spectral matrix  $U$  in the direction
$\xi\in C^{\infty}(\R)$ at point $m$ is
\be
U_{*}(\xi)\stackrel{\triangle}{=}\left.\frac{{\rm d}}{{\rm d} \epsilon}
\right|_{\epsilon=0}U(m+\epsilon\xi)
=\zhen{0}{0}{0}{0}{0}{0}{-\xi\lambda}{0}{0} \label{6.4.5}
\ee
which is obviously an injective homomorphism.
    
   For any given $C^{\infty}$-function $G$, we construct
    the following $3\times 3 $ matrix  equation with respect to
    $V=V(G)$
    \beq   V_x- [U,V]=U_{*}(K G-\lambda^2 J G). \label{VLCH}
    \eeq

  \begin{theorem}
For the spectral problem (\ref{SP1}) and an arbitrary
$C^{\infty}$-function $G$, the matrix equation
(\ref{VLCH})  has the
following solution
{\footnotesize \beq
  V=\lambda
\zhen{\Gamma G+3\lambda\pa\Theta ^{-1}\Upsilon G}{3G_x-3\lambda\Theta ^{-1}\Upsilon G}{-6G}
{\Gamma G_x+3\lambda(\pa^2\Theta ^{-1}\Upsilon G+2mG)}{-2(G-G_{xx})}
{-3G_x-3\lambda\Theta ^{-1}\Upsilon G}
{\Gamma G_{xx}+3\lambda(\pa+\lambda m)\Theta ^{-1}\Upsilon G}
{-\Theta  G-3\lambda(\pa^{-1}\Upsilon G-2mG)}{-2G-G_{xx}-3\lambda\pa\Theta ^{-1}\Upsilon G}\nonumber\\
 \label{6.4.15}
\eeq}
where $\Theta=\pa-\pa^3, \ \Upsilon=m\pa+2\pa m, \ \Gamma=4-\pa^2$. Therefore,
$J=3\Upsilon^*\Theta ^{-1}\Upsilon, \ K=\Gamma\Theta  $.
\label{Th1}
\end{theorem}
{\bf Proof:} \ \ Let 
\beqq
V=\zhen{V_{11}}{V_{12}}{V_{13}}{V_{21}}{V_{22}}{V_{23}}{V_{31}}{V_{32}}{V_{33}},
\eeqq 
and substitute it into Eq. (\ref{VLCH}). That is a over-determined
equation. Using some calculation technique \cite{Qiao}, we obtain the following
results:
\beqq
V_{11}&=&\lambda \Gamma G+3\lambda^2\pa\Theta ^{-1}\Upsilon G,\\
V_{12}&=&  3\lambda G_x-3\lambda^2\Theta ^{-1}\Upsilon G,\\
V_{13}&=& -6\lambda G,\\
V_{21}&=&  \lambda \Gamma G_x+3\lambda^2(\pa^2\Theta ^{-1}\Upsilon G+2mG),\\
V_{22}&=& -2\lambda(G-G_{xx}),\\
V_{23}&=& -3\lambda G_x-3\lambda^2\Theta ^{-1}\Upsilon G,\\
V_{31}&=& \lambda \Gamma G_{xx}+3\lambda^2(\pa+\lambda m)\Theta ^{-1}\Upsilon G,
\\
V_{32}&=& -\lambda \Theta  G-3\lambda^2(\pa^{-1}\Upsilon G-2mG),\\
V_{33}&=& -2\lambda G-\lambda G_{xx}-3\lambda^2\pa\Theta ^{-1}\Upsilon G,
\eeqq
which completes the proof.

\begin{theorem}
Let $G_0\in Ker\ J=\{G\in C^{\infty}(\R)\ |\ JG=0\}$ and
$G_{-1}\in Ker\ K=\{G\in C^{\infty}(\R)\ |\ KG=0\}$. We define
the Lenard's sequences as follows
\beq
G_j= {\cal L}^j G_0 ={\cal L}^{j+1} G_{-1}, \ j\in \Z. \label{Gj}
\eeq
Then, 
\begin{enumerate}
\item 
  the all vector fields $X_k=J G_k, \ k\in \Z$ satisfy the
following commutator representation
\beq 
V_{k,x}-[U,V_k]=U_*(X_k), \ \forall k\in \Z; \label{UV1}
\eeq
\item  the following hierarchy of nonlinear evolution equations
\beq
m_{t_k}= X_k=J G_k, \ \forall k \in \Z, \label{mtk}
\eeq
possesses the zero curvature representation
\beq
U_{t_k}-V_{k,x}+[U,V_k]=0,\ \forall k\in \Z, \label{XUV}
\eeq
\end{enumerate}
where 
\beq
V_k&=&\sum V(G_j)\lambda^{2(k-j-1)}, \ \ \sum\ = \
\left\{\begin{array}{ll}
\sum^{k-1}_{j=0}, & k>0,\\
0, & k=0,\\
-\sum^{-1}_{j=k}, & k<0,
\end{array}\right.  
\eeq
and $V(G_j)$ is given by Eq. (\ref{6.4.15}) with $G=G_j$.
\label{Th12}
\end{theorem}

{\bf Proof:} \begin{enumerate}
\item  For $k=0$, it is obvious.
For $k<0$, we have
\beqq
V_{k,x}-[U,V_k]&=&-\sum_{j=k}^{-1}\left(V_{x}(G_j)-[U,V(G_j)]\right)\lambda^{2(k-j-1)}\\
&=& -\sum_{j=k}^{-1}U_*\left(K G_j-\lambda^2 K G_{j-1}\right)\lambda^{2(k-j-1)}\\
&=& U_*\left(\sum_{j=k}^{-1} K G_{j-1}\lambda^{2(k-j)}-K
G_{j}\lambda^{2(k-j-1)}\right) \\
&=& U_*\left( K G_{k-1}-K
G_{-1}\lambda^{2k}\right) \\
&=& U_*( K G_{k-1})\\
&=& U_*(X_k).
\eeqq 

For the case of $k>0$, it is similar to prove.

\item  Noticing $U_{t_k}=U_*(m_{t_k})$, we obtain
\beqq
U_{t_k}-V_{k,x}+[U,V_k]= U_*(m_{t_k}-X_k).
\eeqq
The injectiveness of $U_*$ implies item 2 holds.

\end{enumerate}

So, the hierarchy  (\ref{mtk}) has  Lax pair and 
is therefore integrable. In particular, 
 through  choosing $G_{-1}=-\frac{1}{6}\in Ker\ K$, 
(\ref{mtk}) reads
\beq
m_{t_k}=-J{\cal L}^{k+1}\cdot  \frac{1}{6}, \ \ k=-1,-2,..., \label{mtk1} 
\eeq
where ${\cal L}=J^{-1}K$. Set $m=u-u_{xx}$, then it is easy to see
the first equation in the hierarchy is exactly the DP equation
\cite{DP[1999]}
\beq
m_t+um_x+3mu_x=0,\  t=t_{-1}. \label{mtd}
\eeq
This equation 
has the following Lax pair:
\beq
\Psi_x&=&U(m,\lambda)\Psi, \nonumber \\
\Psi_t&=&V(m,\lambda)\Psi, \label{Vt}
\eeq
where $U(m,\lambda) $ is defined by Eq. (\ref{U1}),
and $V(m,\lambda)$ is given by
\beq
V(m,\lambda)=\zhen{u_x+\frac{2}{3}\lambda^{-1}}{-u}{-\lambda^{-1}}{u}{-\frac{1}{3}\lambda^{-1}}
{-u}{u_x+um\lambda}{0}{-u_x-\frac{1}{3}\lambda^{-1}},
\eeq
which can be changed to the form in Ref. \cite{DHH2002}
\beq
\psi_t+\lambda^{-1}\psi_{xx}+u\psi_x-\left(u_x+\frac{2}{3}\lambda^{-1}\right)\psi=0,
\ \psi=\psi_1.
\label{time2}
\eeq

Let us choose a kernel element $G_0$ from Ker$J$:
$G_0=m^{-\frac{2}{3}}$. Then Eq. (\ref{mtk}) reads the
following integrable  hierarchy
\beq
m_{t_k}=J{\cal L}^km^{-\frac{2}{3}}, \ k=0,1,2,.... \label{mtk2}
\eeq
In particular, the equation
\beq
m_t=4(m^{-\frac{2}{3}})_x-5(m^{-\frac{2}{3}})_{xxx}+
(m^{-\frac{2}{3}})_{xxxxx}, \label{mt2}
\eeq 
 has the Lax pair:
\beq
\Psi_x&=&U(m,\lambda)\Psi, \nonumber \\
\Psi_t&=&V_1(m,\lambda)\Psi, \label{V1t}
\eeq
where $U(m,\lambda) $ is defined by Eq. (\ref{U1}),
and $V_1(m,\lambda)$ is given by {\footnotesize
\beq
V_1(m,\lambda)=\lambda\zhen{\Gamma m^{-\frac{2}{3}}}
   {3(m^{-\frac{2}{3}})_x}{-6m^{-\frac{2}{3}}}
{\Gamma(m^{-\frac{2}{3}})_x+6\lambda m^{\frac{1}{3}} }
{2(m^{-\frac{2}{3}})_{xx}-2m^{-\frac{2}{3}}}
{-3(m^{-\frac{2}{3}})_x}
{\Gamma (m^{-\frac{2}{3}})_{xx}}
{(m^{-\frac{2}{3}})_{xxx}-(m^{-\frac{2}{3}})_{x}+6\lambda
  m^{\frac{1}{3}}}{-2m^{-\frac{2}{3}}-(m^{-\frac{2}{3}})_{xx}}
\nonumber\\
\label{V1m}
\eeq}
with the operator $\Gamma=4-\pa^2 $.
Eq. (\ref{mt2}) is therefore a new integrable equation. 

In the next two sections we will give
parametric solutions for the negative order
hierarchy (\ref{mtk1}) and the positive order
hierarchy (\ref{mtk2}).

\section{Parametric solution of the negative order hierarchy 
(\ref{mtk1})}

To get the parametric solution,
we use  the  constrained method which leads finite dimensional 
integrable systems to the PDEs.
Becasue Eq. (\ref{SP1}) is a 3rd order eigenvalue problem,
we have to investigate itself together with its adjoint problem when we adopt
the nonlinearized procedure \cite{Cao1989}. Ma and Strampp \cite{MS1}
ever studied the AKNS and its its adjoint problem, a $2\times 2$ case,
by using the so-called symmetry constraint method. 
Now, we are
discussing a $3\times 3 $ problem related to the hierarchy
 (\ref{mtk}). 
 Let us return to  the spectral problem (\ref{SP1})
and consider its adjoint problem
\beq
\Psi^*_x=\zhen{0}{0}{m\lambda}{-1}{0}{-1}{0}{-1}{0}\Psi^*, \ \Psi^*=\phlem{1}{2}{3}, \label{SP2}
\eeq
where $\psi^*=\psi^*_3 $. 
\subsection{Nonlinearized spectral problems on a symplectic submanifold}

Let $\lambda_j\ (j=1,...,N)$ be $N$ distinct spectral values of (\ref{SP1}) and (\ref{SP2}),
and $q_{1j}, q_{2j}, q_{3j}$ and $p_{1j}, p_{2j}, p_{3j}$ be the corresponding
spectral functions, respectively. Then we have
\beq \begin{array}{l}
q_{1x}=q_2,\\
q_{2x}=q_3,\\
q_{3x}=-m\La q_1+q_2, \end{array} \label{q123}
\eeq
and 
\beq
\begin{array}{l}
p_{1x}=m\La p_3,\\
p_{2x}=-p_1-p_3,\\
p_{3x}=-p_2,\end{array}\label{p123}
\eeq
where $\La=diag(\lambda_1,...,\lambda_N)$, 
$q_k=(q_{k1}, q_{k2},..., q_{kN})^T, \ p_k=(p_{k1}, p_{k2},..., p_{kN})^T, \ k=1,2,3.$

Now, we consider the following $(6N-2)$-dimensional symplectic submanifold in $\R^{6N}$:
\beq
M=\{(p,q)^T\in\R^{6N}|\ F=0,\ G=0 \}\label{ssb}
\eeq
where $p=(p_1,p_2,p_3)^T, \ q=(q_1,q_2,q_3)^T$, $F=\left<\La q_1, p_3\right>-1$,
$G=\left<\La q_2, p_3\right>-\left<\La q_3, p_2\right>+\left<\La q_2, p_1\right>$,
and $\left<\cdot, \cdot\right>$ stands for the standard inner product in $\R^N$.

When we restrict the above two systems in $\R^{6N}$ to the submanifold $M$, 
we obtain a constraint of $m$ relationship to the spectral function $p,q$:
\beq
m=2\frac{\left<\La q_2, p_2\right>-\left<\La q_3, p_1+p_3\right>}{\left<\La^2 q_2, p_3\right>+\left<\La^2 q_1, p_2\right>}.\label{mcons1}
\eeq

Notice: $\left<\La^2 q_2, p_3\right>+\left<\La^2 q_1, p_2\right>\not=0$ is necessary because it assures $M$ is a symplectic submanifold in $\R^{6N}$.

Under the constraint (\ref{mcons1}) the two systems (\ref{q123})and (\ref{p123}) are nonlinearized as follows:
\beq
\begin{array}{l}
q_{1x}=q_2,\\
q_{2x}=q_3,\\
q_{3x}=-2\frac{\left<\La q_2, p_2\right>-\left<\La q_3, p_3\right>-\left<\La q_3, p_1\right>}{\left<\La^2 q_2, p_3\right>+\left<\La^2 q_1, p_2\right>}\La q_1+q_2;\end{array} \label{pq1}
\eeq
and 
\beq
\begin{array}{l}
p_{1x}=2\frac{\left<\La q_2, p_2\right>-\left<\La q_3, p_3\right>-\left<\La q_3, p_1\right>}{\left<\La^2 q_2, p_3\right>+\left<\La^2 q_1, p_2\right>}\La p_3,\\
p_{2x}=-p_1-p_3,\\
p_{3x}=-p_2.\end{array} \label{pq2}
\eeq

They are forming a $(6N-2)$ dimensional nonlinear system on $M$ with respect to $p,q$.
Is it integrable? To see this, in $\R^{6N}$ we modify the usual
standard Poisson bracket \cite{AV} of two functions $F_1,F_2$ as follows:
\beq
\{F_1, F_2\}=\sum_{i=1}^3\left(\left<\frac{\pa F_1}{\pa q_i}, \frac{\pa F_2}{\pa p_i}\right>-\left<\frac{\pa F_1}{\pa p_i}, \frac{\pa F_2}{\pa q_i}\right>\right) \label{Poiss1}
\eeq  
which is still antisymmetric, bilinear and satisfies the Jacobi identity.

Obviously, 
\beq
\{F,G\}=\left<\La^2 q_2, p_3\right>+\left<\La^2 q_1, p_2\right>\not=0.
\eeq

Because we are discussing the system on the submanifold $M$,
we need to introduce the so-called Dirac-Poisson bracket of two functions
$f,g$ on $M$:
\beq
\{f,g\}_D=\{f,g\}+\frac{1}{\{F,G\}}\left(\{f,F\}\{G,g\}-\{f,G\}\{F,g\}\right)
\label{Dirac1}\eeq
which is  satisfying the Jacobi identity.

Now, we choose a simple Hamiltonian
\beq
H=\left<q_2,p_1+p_3\right>+\left<q_3,p_2\right>, \label{H1}
\eeq
then, the two systems (\ref{pq1}) and (\ref{pq2}) have the canonical Hamiltonian form on $M$:
\beq \begin{array}{l}
q_{1j,x}=\{q_{1j},H\}_D,\\
q_{2j,x}=\{q_{2j},H\}_D,\\
q_{3j,x}=\{q_{3j},H\}_D;\\
p_{1j,x}=\{p_{1j},H\}_D,\\
p_{2j,x}=\{p_{2j},H\}_D,\\
p_{3j,x}=\{p_{3j},H\}_D,
\end{array}
\eeq
which can be in a brief form rewritten as:
\beq \begin{array}{l}
q_x=\{q, H\}_D,\\
p_x=\{p, H\}_D. \end{array}\label{pqH1}
\eeq
In this calculation procedure, we have used 
\beqq
\{H,G\}&=&2\Big(\left<\La q_2, p_2\right>-\left<\La q_3, p_1+p_3\right>\Big),\\
\{H,F\}&=&0.
\eeqq 

It is easy to check that $H_x=0$, i.e. $H$ is invariant along the flow
(\ref{pqH1}). Assume $H=C_D$ ($C_D$ is constant) along this flow and
\beq
u=\frac{1}{2}\left(\left<q_1,p_2\right>+\left<q_2,p_3\right>\right)-\frac{1}{2}H, \label{u1}
\eeq
then we have
\beq
u-u_{xx}=m,
\eeq
which is exactly related to the DP equation (\ref{mtd}).

To show the integrability of canonical system (\ref{pqH1}),
we need to consider the nonlinearization
of the time part of the Lax representations.

\subsection{Nonlinearized time part on this submanifold}
Let us turn to the time part (\ref{Vt}) of the Lax pair
for the TD equation (\ref{mtd}).
Then the corresponding adjoint problem
 reads
\beq
\Psi^*_t=\zhen{-u_x-\frac{2}{3}\lambda^{-1}}{-u}{-u_x-um\lambda}
{u}{\frac{1}{3}\lambda^{-1}}{0}
{\lambda^{-1}}{u}{u_x+\frac{1}{3}\lambda^{-1}}\Psi^*, \ \Psi^*=\phlem{1}{2}{3}. \label{TSP2}
\eeq
We also consider the constrained system of the time part on $M$.
Thus, under the constraints (\ref{mcons1}), (\ref{u1}), and
\beq
u_x&=&\frac{1}{2}\left(\left<q_3-q_1,p_3\right>-\left<q_1,p_1\right>\right),
\label{ux1}
\eeq
Eqs. (\ref{Vt}) and (\ref{TSP2}) are nonlinearized as:
\beq
\begin{array}{l}
q_{1t}=u_xq_1+\frac{2}{3}\La^{-1}q_1-uq_2-\La^{-1}q_3,\\
q_{2t}=uq_1-\frac{1}{3}\La^{-1}q_2-uq_3,\\
q_{3t}=u_xq_1+um\La q_1-\frac{1}{3}\La^{-1}q_3-u_xq_3;
\end{array} \label{tpq1}
\eeq
and 
\beq
\begin{array}{l}
p_{1t}=-u_xp_1-\frac{2}{3}\La^{-1}p_1-up_2-u_xp_3-um\La p_3,\\
p_{2t}=up_1+\frac{1}{3}\La^{-1}p_2,\\
p_{3t}=\La^{-1}p_1+up_2+u_xp_3+\frac{1}{3}\La^{-1}p_3,
\end{array} \label{tpq2}
\eeq
respectively, where each $q_k,p_k$ and $\La$ are the same as section 2, and
$\La^{-1}$ is the inverse of $\La$.

Let
\beq
I&=&\frac{2}{3}\left<\La^{-1}q_1,p_1\right>-\frac{1}{3}\left<\La^{-1}q_2,p_2\right>
-\frac{1}{3}\left<\La^{-1}q_3,p_3\right>-\left<\La^{-1}q_3,p_1\right>\nonumber\\
& &+\frac{1}{4}\left<q_1,p_2\right>^2-\frac{1}{4}\left<q_2,p_3\right>^2-\frac{1}{2}H\left<q_1,p_2\right> \nonumber\\
& &-u\left(\left<q_2,p_1\right>+\left<q_3,p_2\right>-H\right)-u_x^2,
\eeq
then the two systems (\ref{tpq1}) and (\ref{tpq2}) are  expressed in a canonical
Hamiltonian form on $M$:
\beq \begin{array}{l}
q_t=\{q, I\}_D,\\
p_t=\{p, I\}_D, \end{array}\label{pqI1}
\eeq
where $p=(p_1,p_2,p_3)^T, q=(q_1,q_2,q_3)^T$.

In the above calculations, we use the following equalities:
\beqq
F&=&\left<\La q_1, p_3\right>-1=0,\\
G&=&\left<\La q_2, p_1+p_3\right>-\left<\La q_3, p_2\right>=0,\\
F_x&=&\left<\La q_2, p_3\right>-\left<\La q_1, p_2\right>=0,\\
F_{xx}&=&\left<\La q_3, p_3\right>-2\left<\La q_2, p_2\right>+
\left<\La q_1, p_1\right>+1=0,\\
\{G,I\}&=&2u\Big(\left<\La q_2, p_2\right>-\left<\La q_3, p_1+p_3\right>\Big),\\
\{F,I\}&=&0.
\eeqq 

\begin{theorem}

\beq 
\{H,I\}_D=0,
\eeq
that is, two Hamiltonian flows commute on $M$.
\end{theorem}
{\bf Proof:}\ \ By the definition,
\beq
\{H,I\}_D=\{H,I\}+\frac{1}{\{F,G\}}\left(\{H,F\}\{G,I\}-\{H,G\}\{F,I\}\right),
\label{Dirac2}
\eeq
we need to compute each Poisson bracket in this equality.
\beqq
\{H,I\}&=&u_x\left(\left<p_2,q_1-q_3\right>-\left<q_2,p_1\right>\right)
         +u\left(\left<p_3,q_1-q_3\right>-\left<q_1,p_1\right>\right)=0,\\
\{H,F\}&=&\left<\La q_1,p_2\right>-\left<\La
  q_2,p_3\right>=(-\left<\La q_1,p_3\right>+1)_x=0,\\
\{F,I\}&=&2u_x(\left<\La q_1,p_3\right>-1)+u(-\left<\La q_1,p_3\right>+1)_x=0,
\eeqq
complete the proof.

By Theorem 2, we know that the hierarchy (\ref{mtk1}) has the 
Lax representation:
\beq
\Psi_x&=&U(m,\lambda)\Psi,\\
\Psi_{t_j}&=&-\sum_{k=j}^{-1}V(G_k)\lambda^{2(j-k-1)}\Psi, \ j<0, \label{Vj-}
\eeq
where $V(G_k)$ is given by Eq. (\ref{6.4.15}) with $G=G_k$.

In last section, we have investigated
the nonlinearized systems of spectral problem and its
adjoint. Now, we study the nonlinearizations of time part 
(\ref{Vj-}) and its adjoint problem:
\beq
\Psi^*_{t_j}&=&\sum_{k=j}^{-1}V^T(G_k)\lambda^{2(j-k-1)}\Psi^*, \ j<0, \label{VTj-}
\eeq
where $ V^T(G_k)$ is the transpose of $ V(G_k)$.

Let  $\Psi_k=(q_{1k},q_{2k}, q_{3k} )^T$,
$\Psi^*_k=(p_{1k},p_{2k}, p_{3k} )^T$ be the eigenfunctions
corresponding to $N$ the eigenvalues 
$\lambda_k $ $(k=1,...,N)$ of (\ref{SP1}) and (\ref{SP2}).
 Let us start from
the constraint $G_{-1}=-\frac{1}{6}\sum^N_{j=1}\nabla\lambda_j $.
This constraint is giving the symplectic submanifold $M$ we need. Let
the two antisymmetric operators act on this constraint, we
have:
\beq
G_j=-\frac{1}{6}\left<\La^{2j+3}
      q_1,p_3\right>, \ j<0.
\eeq
Therefore, a complicated calculation yields 
 the following formulations:
\beqq
G_j-G_{j,xx}&=&\frac{1}{6}\Big(\left<\La^{2j+3}
      q_1,p_1\right>+\left<\La^{2j+3}
      q_3,p_3\right>-2\left<\La^{2j+3}
      q_2,p_2\right>\Big),\\
\Gamma G_j&=&\frac{1}{6}\Big(\left<\La^{2j+3}
      q_1,p_1\right>+\left<\La^{2j+3}
      q_3,p_3\right>-2\left<\La^{2j+3}
      q_2,p_2\right>-3\left<\La^{2j+3}
      q_1,p_3\right>\Big),\\
\Gamma G_{j,x}&=&\frac{1}{2}\Big(\left<\La^{2j+3}
      q_1,p_2\right>+\left<\La^{2j+3}
      q_2,p_1\right>-\left<\La^{2j+3}
      q_3,p_2\right>\Big),\\
\Gamma G_{j,xx}&=&\frac{1}{2}\Big[m\Big(\left<\La^{2j+4}
      q_1,p_2\right>+\left<\La^{2j+4}
      q_2,p_3\right>\Big)\\ & &
      +2\left<\La^{2j+3}
      q_3,p_1\right>+\left<\La^{2j+3}
      q_3,p_3\right>-\left<\La^{2j+3}
      q_1,p_1+p_3\right>\Big],\\
\Theta G_{j}&=&\frac{1}{2}\Big(\left<\La^{2j+3}
      q_2,p_1+p_3\right>-\left<\La^{2j+3}
      q_3,p_2\right>\Big),\\ 
\pa^{-1}\Upsilon G_j-2mG_j&=&-\frac{1}{6}\Big(\left<\La^{2j+2}
      q_2,p_1+p_3\right>+\left<\La^{2j+2}
      q_3,p_2\right>\Big),\\
\pa^2\Theta ^{-1}\Upsilon G_j+2mG_j&=&\frac{1}{6}\Big(\left<\La^{2j+2}
      q_2,p_1\right>+\left<\La^{2j+2}
      q_3,p_2\right>-\left<\La^{2j+2}
      q_1,p_2\right>\Big),\\
\Theta ^{-1}\Upsilon G_j&=&-\frac{1}{6}\Big(\left<\La^{2j+2}
      q_1,p_2\right>+\left<\La^{2j+2}
      q_2,p_3\right>\Big).
\eeqq

Substituting these equalities 
to Eqs. (\ref{Vj-}) and (\ref{VTj-}),
with a similar computational  method to section 3 we find
the nonlinearized systems of the time part
(\ref{Vj-}) and the adjoint time part (\ref{VTj-})
are cast in a canonical Hamiltonian system on the $(6N-2)$-dimensional 
symplectic submanifold $M$:
\beq\begin{array}{l}
q_{t_k}=\{q, F_k\}_D,\\
p_{t_k}=\{p, F_k\}_D,\end{array} \ \ \  k=-1,-2,-3,...,\label{pqFk}
\eeq
where
{\scriptsize
\beqq
F_k&=&
\frac{2}{3}\left<\La^{2k+1}q_1,p_1\right>
-\frac{1}{3}\left<\La^{2k+1}q_2,p_2\right>
-\frac{1}{3}\left<\La^{2k+1}q_3,p_3\right>
-\left<\La^{2k+1}q_3,p_1\right>\\
& & 
+\sum_{j=k}^{-2}\Big[
-\frac{1}{12}\left<\La^{2j+3}q_1,p_1\right>\left<\La^{2(k-j)-1}q_1,p_1\right>-\frac{1}{3}\left<\La^{2j+3}q_2,p_2\right>\left<\La^{2(k-j)-1}q_2,p_2\right>-\frac{1}{4}\left<\La^{2j+3}q_3,p_3\right>\left<\La^{2(k-j)-1}q_3,p_3\right>\\
& &
+\frac{1}{3}\left<\La^{2j+3}q_2,p_2\right>\left<\La^{2(k-j)-1}q_1,p_1\right>
-\frac{1}{6}\left<\La^{2j+3}q_3,p_3\right>\left<\La^{2(k-j)-1}q_1,p_1\right>
+\frac{1}{3}\left<\La^{2j+3}q_3,p_3\right>\left<\La^{2(k-j)-1}q_2,p_2\right> \\
& &-\frac{1}{4}\left<\La^{2j+3}q_1,p_2\right>\left<\La^{2(k-j)-1}q_1,p_2\right>+\frac{1}{4}\left<\La^{2j+3}q_1,p_3\right>\left<\La^{2(k-j)-1}q_1,p_3\right>+\frac{1}{4}\left<\La^{2j+3}q_2,p_3\right>\left<\La^{2(k-j)-1}q_2,p_3\right>\\
& &
+\frac{1}{2}\Big(\left<\La^{2j+3}q_1,p_2\right>-\left<\La^{2j+3}q_2,p_3\right>\Big)\Big(\left<\La^{2(k-j)-1}q_3,p_2\right>-\left<\La^{2(k-j)-1}q_2,p_1\right>\Big)\\
& &+\frac{1}{2}\left<\La^{2j+3}q_1,p_3\right>
\Big(\left<\La^{2(k-j)-1}q_1,p_1\right>-2\left<\La^{2(k-j)-1}q_3,p_1\right>-\left<\La^{2(k-j)-1}q_3,p_3\right>\Big)\Big]\\
& & +\sum_{j=k}^{-1}\Big[
-\frac{1}{4}\left<\La^{2j+2}q_1,p_1\right>\left<\La^{2(k-j)}q_1,p_1\right>-\frac{1}{4}\left<\La^{2j+2}q_3,p_3\right>\left<\La^{2(k-j)}q_3,p_3\right>\\
& &
+\frac{1}{4}\left<\La^{2j+2}q_1,p_2\right>\left<\La^{2(k-j)}q_1,p_2\right>-\frac{1}{4}\left<\La^{2j+2}q_1,p_3\right>\left<\La^{2(k-j)}q_1,p_3\right>-\frac{1}{4}\left<\La^{2j+2}q_2,p_3\right>\left<\La^{2(k-j)}q_2,p_3\right>\\
& & 
-\frac{1}{2}\Big(\left<\La^{2j+2}q_1,p_2\right>+\left<\La^{2j+2}q_2,p_3\right>\Big)\Big(\left<\La^{2(k-j)}q_3,p_2\right>+\left<\La^{2(k-j)}q_2,p_1\right>\Big)\\
& &+\frac{1}{2}\left<\La^{2j+2}q_1,p_3\right>
\left<\La^{2(k-j)}q_3,p_3\right>+\frac{1}{2}\left<\La^{2j+2}q_3,p_3\right>\left<\La^{2(k-j)}q_1,p_1\right>-\frac{1}{2}\left<\La^{2j+2}q_1,p_3\right>\left<\La^{2(k-j)}q_1,p_1\right>\Big].
\eeqq}

Apparently, $F_{-1}=I $. Furthermore, by the Dirac-Poisson bracket (\ref{Dirac1}) on
submanifold $M$ we obtain the following theorem.
\begin{theorem}
All Hamiltonian flows (\ref{pqH1}) and
(\ref{pqFk}) commute on $M$.
\label{TH4}
\end{theorem}
{\bf Proof}: \ \  Through a lengthy calculation,
we have
\beqq
\{H, F_k\}_D=0,
\eeqq
which completes the proof.

$ $

Therefore, all Hamiltonian flows (\ref{pqFk}) 
are integrable on $M$. Particularly,  the Hamiltonian system $(\ref{pqH1})$
is integrable.

\begin{remark}  In the calculation procedure
and the proof of Theorem \ref{TH4} and Eq. (\ref{pqFk}), 
we used the following facts:
\beqq
& & \{q_1,F\}=\{q_2,F\}=0,\\
& & \{q_3, F\}=\La q_1,\\
& & \{p_1,F\}=\{p_2,F\}=0,\\
& & \{p_3, F\}=-\La p_3,\\
& & \{F,F_k\}=0,\\
& & \{G,F_k\}=(\left<\La q_2,p_2\right>-\left<\La q_3,p_1+p_3\right>)(\left<\La^{2k+2}q_1,p_2\right>+\left<\La^{2k+2}q_2,p_3\right>)  ,
\eeqq
and on $M$ all of these equalities $F=G=F_x=F_{xx}=0$ are valid.
\end{remark}

\subsection{Parametric solution}

\begin{theorem}
Let $p(x,t_k),q(x,t_k)$ ($p(x,t_k)=(p_1, p_2,p_3)^T,  q(x,t_k)=(q_1,
q_2,q_3)^T, \ k=-1,-2,...$) be the common solution of the two integrable flows
(\ref{pqH1}) and (\ref{pqFk}), then 
\beq
m&=&2\frac{\left<\La q_2, p_2\right>-\left<\La q_3, p_1+p_3\right>}{\left<\La^2 q_2, p_3\right>+\left<\La^2 q_1, p_2\right>},\label{mcons-general}
\eeq
satisfies the  negative order hierarchy (\ref{mtk1}).
\end{theorem}
{\bf Proof:} \ \ Noticing the following formulas 
\beqq
G_k&=&-\frac{1}{6}\left<\La^{2k+3}
      q_1,p_3\right>, \ k=-1,-2,-3,...,\\
\Theta ^{-1}\Upsilon G_k&=&-\frac{1}{6}\Big(\left<\La^{2k+2}
      q_1,p_2\right>+\left<\La^{2k+2}
      q_2,p_3\right>\Big),\\
\Upsilon&=&m\pa  +2\pa m, \ \ \Upsilon^*=\pa m +2m\pa, \ \ \Theta=\pa-\pa^3,
\eeqq
and Eq. (\ref{pqFk}), we 
directly compute and find  Eq. (\ref{mcons-general})
satisfies $m_{t_k}=3\Upsilon^*\Theta ^{-1}\Upsilon G_k $
which completes the proof.  

$ $

In particular, we obtain the following theorem.

\begin{theorem}
Let $p(x,t),q(x,t)$ ($p(x,t)=(p_1, p_2,p_3)^T,  q(x,t)=(q_1, q_2,q_3)^T$) be the common solution of the two integrable flows
(\ref{pqH1}) and (\ref{pqI1}), then 
\beq
m&=&2\frac{\left<\La q_2(x,t), p_2(x,t)\right>-\left<\La q_3(x,t), p_1(x,t)+p_3(x,t)\right>}{\left<\La^2 q_2(x,t), p_3(x,t)\right>+\left<\La^2 q_1(x,t), p_2(x,t)\right>},\label{mcons3}\\
u&=&\frac{1}{2}\left(\left<q_1(x,t),p_2(x,t)\right>+\left<q_2(x,t),p_3(x,t)\right>\right)-\frac{1}{2}H, \label{u3}
\eeq
satisfy the DP equation:
\beq
m_t+um_x+3mu_x=0.
\eeq

\end{theorem}
{\bf Proof:} \ \ Let 
\beqq
A&=&\left<\La q_2, p_2\right>-\left<\La q_3, p_1+ p_3\right>,\\
B&=&\left<\La^2 q_2, p_3\right>+\left<\La^2 q_1, p_2\right>,\\
C&=&\left<\La^2 q_3, p_3\right>-\left<\La^2 q_1, p_1+ p_3\right>.
\eeqq
Then through a lengthy calculation we have
\beqq
A_t&=&uG+umC-2u_xA=umC-2u_xA,\\
B_t&=&G-uC+u_xB=-uC+u_xB,\\
A_x&=&-2G-mC=-mC,\\
B_x&=&C.
\eeqq

By the above equalities, we obtain
\beqq
& &m_t+um_x+3mu_x \\
& &=\frac{2}{B^2}\Big[A_tB-AB_t+u(A_xB-AB_x)+3u_xAB\Big]\\
& &=\frac{2}{B^2}\Big(umBC+uAC+uA_xB-uAB_x\Big)\\
& &=\frac{2u}{B^2}\Big(B(mC+A_x)+A(C-B_x)\Big)\\
& &=0. 
\eeqq
In the above proof procedures, we have used the following equalities:
$F=G=0, \ F_x=F_{xx}=0$ on $M$.

$ $

Similarly, we can discuss the parametric solution of
the positive order hierarchy (\ref{mtk2}).
 That needs us to consider
a new kind of constraint and related integrable system,
which we deal with in the next section.

\section{Parametric solution of
the positive order hierarchy (\ref{mtk2})}

Let us directly consider the following constraint:
\beq
G_0=\sum^N_{j=1}E_j\nabla\lambda_j,\ \label{G0} 
\eeq
where $E_j\nabla\lambda_j=\lambda_jq_{1j}p_{3j}$ is the functional gradient
of $\lambda_j$ for the spectral problems (\ref{SP1}) and (\ref{SP2}), and 
 $q_{kj}, \ p_{kj} \ (k=1,2,3)$ are the related eigenfunctions of
$ \lambda_j$. Then Eq. (\ref{G0}) is saying
\beq
m=\left<\La q_1, p_3\right>^{-\frac{3}{2}} \label{m+}
\eeq
which composes a new constraint in the whole space $\R^{6N}$.
Under this constraint, the spectral problem (\ref{SP1}) and its
adjoint
problem (\ref{SP2})
are able to be cast in a Hamiltonian canonical form in $\R^{6N} $:
\beq 
(H^+): \ \ \ 
\begin{array}{l}
q_x=\{q, H^+\},\\
p_x=\{p, H^+\}, \end{array}\label{pqH+}
\eeq
with the Hamiltonian
\beq
H^+=\left<q_2,p_1+p_3\right>+\left<q_3,p_2\right>+\frac{2}{
\sqrt{\left<\La q_1,p_3\right>}}. \label{H+}
\eeq
To see the integrability of the system (\ref{pqH+}), we
take into account of the time part $\Psi_t=V_1(m,\lambda)\Psi $
and its adjoint $\Psi_t=-V^T_1(m,\lambda)\Psi $, where $V_1(m,\lambda) $
is defined by Eq. (\ref{V1m}). Under the constraint (\ref{m+}), 
the time part and its adjoint are also nonlinearized as
a canonical Hamiltonian system in $\R^{6N} $
\beq
 (F_1): \ \ \ 
\begin{array}{l}
q_t=\{q, F_1\},\\
p_t=\{p, F_1\}, \end{array}\label{pqF1}
\eeq
with the Hamiltonian 
\newpage
\beq
F_1&=& 6\frac{\left<\La^2 q_1,p_2\right>+\left<\La^2 q_2,p_3\right>
}{\sqrt{\left<\La q_1,p_3\right>}}\nonumber\\
& &
-\frac{1}{2}(\left<\La q_1,p_1\right>^2+4\left<\La q_2,p_2\right>^2+
\left<\La q_3,p_3\right>^2)\nonumber\\
& &+\frac{3}{2}(\left<\La q_1,p_3\right>^2+\left<\La q_2,p_3\right>^2-
\left<\La q_1,p_2\right>^2)\nonumber\\
& &+3\left<\La q_1,p_3\right>(\left<\La q_1,p_1\right>-2\left<\La q_3,p_1\right>-
\left<\La q_3,p_3\right>)\nonumber\\
& &+2\left<\La q_1,p_1\right>\left<\La q_2,p_2\right>+
2\left<\La q_2,p_2\right>\left<\La q_3,p_3\right>-
\left<\La q_1,p_1\right>\left<\La q_3,p_3\right>\nonumber\\
& &+3(\left<\La q_1,p_2\right>-\left<\La q_2,p_3\right>)
(\left<\La q_3,p_2\right>-\left<\La q_2,p_1\right>).
 \label{F1}
\eeq
  
After a calculation of the Poisson bracket
of $\{H^+,F_1\} $, we know that the two canonical Hamiltonian flows
commute, i.e. 
\beqq
\{H^+,F_1\}=0.
\eeqq
Furthermore, under the constraint (\ref{m+}) the nonlinearized 
systems of the general time part
$\Psi_{t_k}=\sum_{j=0}^{k-1}V(G_j)\lambda^{2(k-j-1)}\Psi $ ($k>0, \ k
\in \Z$) and its adjoint problem produce the following
canonical Hamiltonian system in  $\R^{6N}$:
\beq (F_k): \ \ \ 
\begin{array}{l}
q_{t_k}=\{q, F_k\},\\
p_{t_k}=\{p, F_k\}, \end{array} k=1,2,3,...,\label{pqFk+}
\eeq
with the Hamiltonian

{\scriptsize
\beqq
F_k&=&6\frac{\left<\La^{2k}q_1,p_2\right>+\left<\La^{2k}q_2,p_3\right>
}{\sqrt{\left<\La q_1,p_3\right>
}}\\
& &+\sum_{j=0}^{k-1}\Big[
-\frac{1}{2}\left<\La^{2j+1}q_1,p_1\right>\left<\La^{2(k-j)-1}q_1,p_1\right>-2\left<\La^{2j+1}q_2,p_2\right>\left<\La^{2(k-j)-1}q_2,p_2\right>-\frac{1}{2}\left<\La^{2j+1}q_3,p_3\right>\left<\La^{2(k-j)-1}q_3,p_3\right>\\
& &
+2\left<\La^{2j+1}q_2,p_2\right>\left<\La^{2(k-j)-1}q_1,p_1\right>
-\left<\La^{2j+1}q_3,p_3\right>\left<\La^{2(k-j)-1}q_1,p_1\right>
+2\left<\La^{2j+1}q_3,p_3\right>\left<\La^{2(k-j)-1}q_2,p_2\right> \\
& &-\frac{3}{2}\left<\La^{2j+1}q_1,p_2\right>\left<\La^{2(k-j)-1}q_1,p_2\right>+\frac{3}{2}\left<\La^{2j+1}q_1,p_3\right>\left<\La^{2(k-j)-1}q_1,p_3\right>+\frac{3}{2}\left<\La^{2j+1}q_2,p_3\right>\left<\La^{2(k-j)-1}q_2,p_3\right>\\
& & +3\Big(\left<\La^{2j+1}q_1,p_2\right>-\left<\La^{2j+1}q_2,p_3\right>\Big)\Big(\left<\La^{2(k-j)-1}q_3,p_2\right>-\left<\La^{2(k-j)-1}q_2,p_1\right>\Big)\\
& &+3\left<\La^{2j+1}q_1,p_3\right>
\Big(\left<\La^{2(k-j)-1}q_1,p_1\right>-2\left<\La^{2(k-j)-1}q_3,p_1\right>-\left<\La^{2(k-j)-1}q_3,p_3\right>\Big)\Big]\\
& & +\sum_{j=1}^{k-1}\Big[
-\frac{3}{2}\left<\La^{2j}q_1,p_1\right>\left<\La^{2(k-j)}q_1,p_1\right>-\frac{3}{2}\left<\La^{2j}q_3,p_3\right>\left<\La^{2(k-j)}q_3,p_3\right>\\
& &
+\frac{3}{2}\left<\La^{2j}q_1,p_2\right>\left<\La^{2(k-j)}q_1,p_2\right>-\frac{3}{2}\left<\La^{2j}q_1,p_3\right>\left<\La^{2(k-j)}q_1,p_3\right>-\frac{3}{2}\left<\La^{2j}q_2,p_3\right>\left<\La^{2(k-j)}q_2,p_3\right>\\
& & 
-3\Big(\left<\La^{2j}q_1,p_2\right>+\left<\La^{2j}q_2,p_3\right>\Big)\Big(\left<\La^{2(k-j)}q_3,p_2\right>+\left<\La^{2(k-j)}q_2,p_1\right>\Big)\\
& &+3\left<\La^{2j}q_1,p_3\right>
\left<\La^{2(k-j)}q_3,p_3\right>+3\left<\La^{2j}q_3,p_3\right>\left<\La^{2(k-j)}q_1,p_1\right>-3\left<\La^{2j}q_1,p_3\right>\left<\La^{2(k-j)}q_1,p_1\right>\Big].
\eeqq
}
$ $\\
Apparently, when $k=1$, $F_k$ is exactly Eq. (\ref{F1}). Furthermore,
through a lengthy computation, we obtain
\beq
\{H^+, F_k\}=0, \ \ k=1,2,...,
\eeq
which represents each Hamiltonian $t$-flow $(F_k)$ commutes with
Hamiltonian $x$-flow $(H^+)$. Thus, all Hamiltonian canonical systems
$(F_k) $ are integrable in $\R^{6N}$. Particularly, the nonlinearized
spectral problems (\ref{pqH+}) is integrable.

Like last section, we also have a similar theorem
 
\begin{theorem}
Let $p(x,t_k),q(x,t_k)$ ($p(x,t_k)=(p_1, p_2,p_3)^T,  q(x,t_k)=(q_1,
q_2,q_3)^T, \ k=1,2,3,...$) be the common solution of the two integrable flows
(\ref{pqH+}) and (\ref{pqFk+}), then 
\beq
m&=&\frac{1}{\sqrt{\left<\La q_1(x,t_k), p_3(x,t_k)\right>^3}}, \ \ k=1,2,3,...,
\eeq
satisfy the  positive order hierarchy (\ref{mtk2}).
\end{theorem}

In particular, we have the following theorem.

\begin{theorem}
Let $p(x,t),q(x,t)$ ($p(x,t)=(p_1, p_2,p_3)^T,  q(x,t)=(q_1, q_2,q_3)^T$) be the common solution of the two integrable flows
(\ref{pqH+}) and (\ref{pqF1}), then 
\beq
m&=&\frac{1}{\sqrt{\left<\La q_1(x,t), p_3(x,t)\right>^3}} 
\label{mps}
\eeq
is a parametric solution of the 5th-order equation:
\beq
m_t=4(m^{-\frac{2}{3}})_x-5(m^{-\frac{2}{3}})_{xxx}+
(m^{-\frac{2}{3}})_{xxxxx}.
\eeq
 \end{theorem}
{\bf Proof}: \ \ A direct check is done
through substitution of Eqs. (\ref{pqH+}) and (\ref{pqF1}).

\section{Peaked stationary solutions}

We know that the equation $m_t+um_x+3mu_x=0, \ m=u-m_{xx}$
has peakon solution $ u=e^{-|x+t|}$. Furthermore, a generalized
$b$-balanced equation $m_t+um_x+bmu_x=0, \ m=u-m_{xx}, b=constant$
is found to have this kind of solution  
 \cite{Holm-Staley}.

Now, we study the traveling wave solution of the equation
 $
m_t=4(m^{-\frac{2}{3}})_x-5(m^{-\frac{2}{3}})_{xxx}+
(m^{-\frac{2}{3}})_{xxxxx}$. Set $m^{-\frac{2}{3}}=v$, then
this equation becomes
\beq
-\frac{3}{2}v^{-\frac{2}{3}}v_t=4v_x-5v_{xxx}+v_{xxxxx}.
\label{v5th-order}
\eeq
Assume this equation has the solution  
 $v=f(\xi), \ \xi=x-ct, \ c=constant $, then we have
\beq
2cf^{-\frac{1}{2}}=2f^2-\frac{5}{2}f'^2+f'f'''-\frac{1}{2}f''^2. \label{fode}
\eeq
The right hand side of this equation is quadratically homogeneous
and the left hand side not. Therefore, we set $f=e^{a\xi}, \ a=constant$
 and substitute it into Eq. (\ref{fode}), and obtain
\beq
c&=&0,\\
a^4-5a^2+4&=&0,
\eeq
which implies 
\beq
a=\pm1, \ \ a=\pm2.
\eeq

So, the 5th-order equation (\ref{v5th-order}) has the following
stationary solutions:
\beq
1, \ e^{-x}, \ e^{x}, \ e^{-2x}, \ e^{2x}, 
\eeq
which exactly composes the basis of the solution space of the
stationary equation $4v_x-5v_{xxx}+v_{xxxxx}=0$. Therefore, 
the 5th order PDE  $
m_t=4(m^{-\frac{2}{3}})_x-5(m^{-\frac{2}{3}})_{xxx}+
(m^{-\frac{2}{3}})_{xxxxx}$ possesses the stationary solutions
\beq
m(x)&=&(c_0+c_1e^{-x}+c_2e^{x}+c_3e^{-2x}+c_4e^{2x})^{-\frac{3}{2}},\label{m-stationary}\\
& & \forall \ c_k\in \R, \ k=0,1,2,3,4. \nonumber
\eeq
Apparently, $e^{-\frac{3}{2}x}$, $e^{\frac{3}{2}x}$,
$e^{-3x}$, $e^{3x}$ satisfy the 5th order PDE  $
m_t=4(m^{-\frac{2}{3}})_x-5(m^{-\frac{2}{3}})_{xxx}+
(m^{-\frac{2}{3}})_{xxxxx}$. Therefore, 
 $f_1^\wedge(x)=e^{-\frac{3}{2}|x|}$, $f_1^\vee(x)=e^{\frac{3}{2}|x|}$,  
 $f_2^\wedge(x)=e^{-3|x|}$ , and $f_2^\vee(x)=e^{3|x|}$
are the four peaked stationary solutions (see Figure 1) of the 5th-order PDE
$
m_t=4(m^{-\frac{2}{3}})_x-5(m^{-\frac{2}{3}})_{xxx}+
(m^{-\frac{2}{3}})_{xxxxx}$.

\begin{figure}[ht!]
\centerline{
\scalebox{0.82}{
\includegraphics{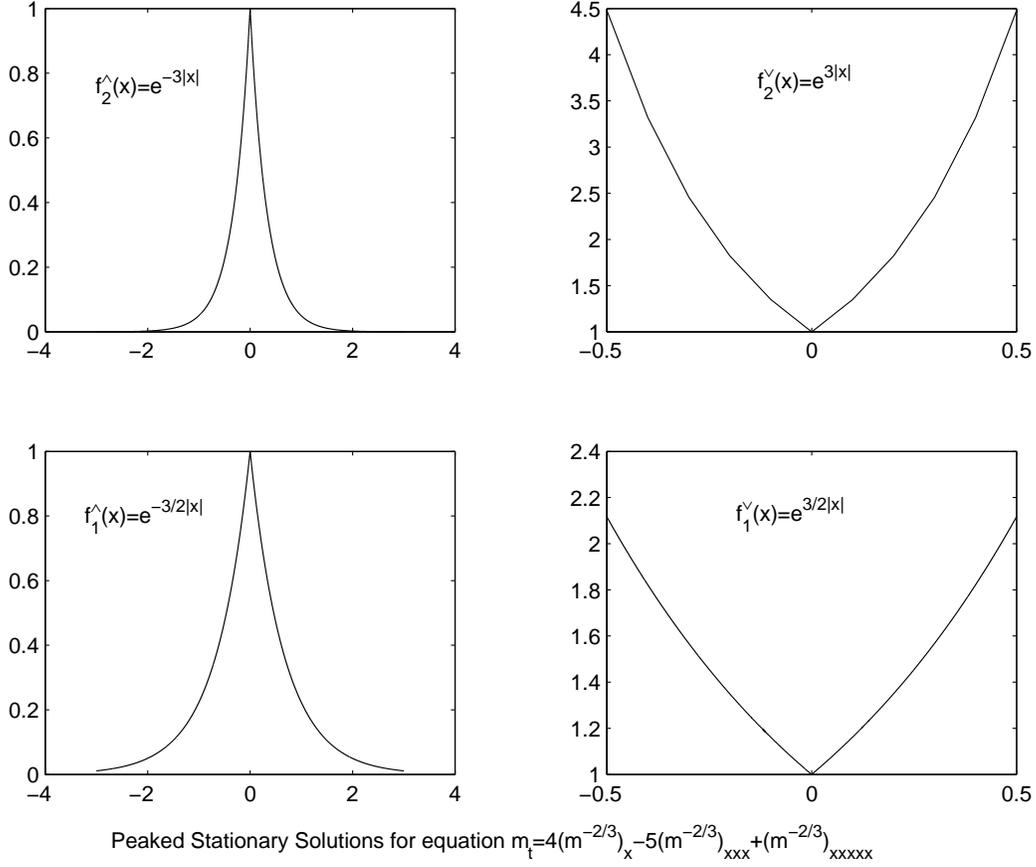}
}
}
\caption{Four peaked stationary solutions}
\label{5thorder-pde-fig}
\end{figure}

\subsection*{Comparison with the parametric solution (\ref{mps})}
All the stationary solutions (\ref{m-stationary}) may be included in the
parametric solution (\ref{mps}). For example,
the function $m(x)=e^{-\frac{3}{2}x}$
is cast in Eq. (\ref{mps}) when we choose dynamical
variables $q_1, \ p_3$ such that the following constraint 
\beq
\left<\La q_1(x), p_3(x)\right>=e^{x}
\eeq
holds, where $q_1, \ p_3 $ are the solutions of the integrable
Hamiltonian system (\ref{pqH+}). But the four peaked stationary
solutions can not be included in the
parametric solution (\ref{mps}) because $\left<\La q_1(x),
  p_3(x)\right>$ is smooth every where. 

\subsection*{Comparison with the cusp solution \cite{Wadati}}

Wadati,  Ichikawa and Shimizu \cite{Wadati} proposed the cusp soliton 
(cuspon) solution for the deformed Harry-Dym equation
$r_t+(1-r)^3r_{xxx}=0$. This equation is actually equivalent to the
Harry-Dym equation $q_t-2(q^{-1/2})_{xxx}=0$ by the transformation
$q^{-1/2}=1-r$. By the inverse scattering method
\cite{AKNS1}, they obtained the following traveling wave solution 
\beq
r=\cosh^{-2}\xi, \ \xi=ax-4a^3t+\tanh\xi+A,\label{r-cusp} 
\eeq 
where $a\not=0, a, \ A=constants $. 
They called this solution the cusp soliton. It seems that the
present 
5th-order PDE $
m_t=4(m^{-\frac{2}{3}})_x-5(m^{-\frac{2}{3}})_{xxx}+
(m^{-\frac{2}{3}})_{xxxxx}$   looks like a higher-order Harry-Dym
equation and should have the cusp-kind solution (\ref{r-cusp}). But
unfortunately, this is not the case for our 5th-order PDE $
m_t=4(m^{-\frac{2}{3}})_x-5(m^{-\frac{2}{3}})_{xxx}+
(m^{-\frac{2}{3}})_{xxxxx}$. However, our recent study \cite{HQ-cusp} reveals that the following 5th-order
PDE
\beq
(15r^2+18r+2)r_t+(1-r)^6r_{xxxxx}=0
\eeq 
possesses the cuspon $r=\cosh^{-2}\xi$ with
$\xi=ax-8a^5t+\tanh\xi+A$, $\forall  a\not=0, A \in \R$.

\subsection*{Acknowledgments}
We would like to express our sincere thanks to Prof. Konopelchenko for his suggestion and Prof. Magri
for his fruitful discussion during their visit at Los Alamos National
Laboratory.

This work was supported by the U.S. Department of Energy under
contracts W-7405-ENG-36 and the Applied Mathematical Sciences Program
KC-07-01-01; and also the Special Grant of National
Excellent Doctorial Dissertation of PR China.


\begin{thebibliography}{99}
\parskip=0.02cm
\small
\bibitem{AKNS}Ablowitz M J,  Kaup D J,  Newell A C,  Segur H,
             Nolinear evolution equations of physical significance,
             {\it Phys. Rev. Lett.}  31(1973), 125-127.

\bibitem{AKNS1}Ablowitz M J,  Kaup D J,  Newell A C,  Segur H,
  {\it Studies in Appl. Math. } 53(1974), 249-315.
\bibitem{AV}V. I. Arnol'd, {\it Mathematical Methods of Classical Mechanics}
            (Springer-Verlag, Berlin, 1978).
\bibitem{CH1}Camassa R, Holm D D,
          An integrable shallow water
          equation with peaked solitons,
                     {\it Phys. Rev. Lett.}  71 (1993), 1661-1664.
\bibitem{Cao1989}Cao C W,
Nonlinearization of Lax system
             for the AKNS hierarchy,  {\it Sci. China A} (in Chinese) 32(1989), 701-707;
             also see English Edition:
              Nonlinearization of Lax system for the AKNS hierarchy,
              {\it Sci. Sin. A} 33(1990), 528-536.
\bibitem{DHH2002}
Degasperis A, Holm D D,  Hone A N W,
A new integrable equation with peakon
solutions, {\it NEEDS (2002) Proceedings}, to appear.

\bibitem{DP[1999]}
Degasperis A and Procesi M, Asymptotic integrability, in Symmetry and Perturbation Theory, edited by A. Degasperis and G. Gaeta, World Scientific (1999) pp.23-37.


\bibitem{GGKM}Gardner C S, Greene J M, Kruskal M D, Miura R M,
         Method for Solving the Korteweg-de Vries Equation, {\it Phys. Rev. Lett.}
         19(1967), 1095-1097.

\bibitem{HH[2002]}Holm D D,  Hone A N W, Note on Peakon Bracket,
Private communication, 2002.
\bibitem{HQ-cusp}Holm D D, Qiao Z, Equations possessing cusp solitons and cusp-like singular traveling  wave  solutions, in
  preparation, 2002.

\bibitem{HQ-new-twv}Holm D D, Qiao Z, 
An integrable hierarchy, parametric solution and traveling wave
solution, preprint, arXiv:nlin.SI/0209026, 2002.


\bibitem{Holm-Staley}Holm D D,  Staley M, Private communication, 2002.


\bibitem{Kaup1}Kaup D J, On the inverse scattering problem for
         cubic eigenvalue problems of the class
         $\psi_{xxx}+6Q\psi_x+6R\psi=\lambda\psi$, {\it
         Stud. Appl. Math.} 62(1980), 189-216.
\bibitem{KD1}Konopelchenko B G,  Dubrovsky V G, Some new
integrable nonlinear evolution equations in $2+1$ dimensions,
 {\it Phys. Lett. A} 102(1984), 15-17.
\bibitem{Kup1}Kuperschmidt B A, A super Korteweg-De Vries
         equation: an integrable system,
 {\it Phys. Lett. A} 102(1984), 213-215.
 
\bibitem{KDV}Korteweg D J,  Vries De G, On the change of form
  long waves advancing in a rectangular canal, and on a new type of
  long stationary waves, {\it Phil. Mag.} 39(1895), 422-443.
\bibitem{LG}Levitan B M, Gasymov M G, Determination of a differential
  equation by two of its spectra, {\it Russ. Math. Surveys} 19:2(1964), 1-63. 
\bibitem{MS1}Ma W X, Strampp W, An explicit symmetry constraint for the Lax pairs and the adjoint Lax pairs of AKNS systems, 
{\it Phys. Lett. A} 185(1994), 277-286. 

\bibitem{Mar}Marchenko V A,  Certain problems in the theory of
  second-order differential operators, {\it Doklady Akad. Nauk SSSR} 72(1950), 457--460 (Russian).

\bibitem{Qiao} Qiao Z,  {\it Finite-dimensional Integrable System and Nonlinear Evolution Equations}, Higher Education
  Press, PR China, 2002.

\bibitem{Tu1990}Tu G Z,  An extension of a theorem on gradients of conserved
            densities of integrable systems, {\it Northeast. Math. J.} 6(1990),
            26-32.
\bibitem{Wadati}Wadati M,  Ichikawa Y H, Shimizu T, Cusp soliton of a new integrable nonlinear
evolution equation, 
{\it Prog. Theor. Phys.} 64(1980), 1959-1967.
\bibitem{ZS}Zakharov V E,  Shabat A B, Exact theory of two dimensional
         self focusing and one dimensional self modulation of waves in nonlinear media,
          {\it Sov. Phys. JETP} 34(1972), 62-69.
\end{thebibliography}
\end{document}